\documentstyle[pre,aps]{revtex}
\begin{document}
\draft
\font\sqi=cmssq8
\def\DR{\rm I\kern-1.45pt\rm R}
\def\DC{\kern2pt {\hbox{\sqi I}}\kern-4.2pt\rm C}
\def\DH{\rm I\kern-1.5pt\rm H\kern-1.5pt\rm I}
\newcommand{\bs}{\mbox{\boldmath $\sigma$}}
\def\theequation{\arabic{equation}}
\twocolumn[\hsize\textwidth\columnwidth\hsize\csname
@twocolumnfalse\endcsname

\title{Relation of the oscillator and Coulomb systems on
spheres and pseudospheres}
\author{Armen Nersessian$^{1,2}$, George Pogosyan$^{1,2}$}
\address{$^1$Bogoliubov Laboratory of Theoretical Physics,
Joint Institute for Nuclear Research,
141980 Dubna, Russia\\
$^2$Department of Theoretical Physics and
International Center for Advanced Studies,\\
Yerevan State University, A. Manoogian St., 1, Yerevan, 375025 Armenia}
\date{\today}
\maketitle
\begin{abstract}
\noindent
It is shown, that  oscillators on  the sphere and the pseudosphere
are related,   by the so-called Bohlin transformation,
with the Coulomb systems on the pseudosphere.
The even states of an oscillator yield
 the conventional Coulomb system on the pseudosphere,
 while the odd states yield the
Coulomb system on the pseudosphere in the presence of magnetic flux tube
 generating  spin $1/2$.
A similar relation  is established
for  the oscillator  on  the (pseudo)sphere
specified by the presence of constant uniform   magnetic field $B_0$
 and the   Coulomb-like system on pseudosphere
 specified by the presence of the magnetic field
 $\frac{B}{2r_0}(|\frac{x_3}{{\bf x}}|-\epsilon)$.
The correspondence between the oscillator and the Coulomb systems
 the higher dimensions is also discussed.
\end{abstract}
]
\noindent
The ($d-$dimensional) oscillator and Coulomb systems
are the most known representatives
of mechanical systems with hidden symmetries which  define the
 $su(d)$ symmetry algebra for the oscillator, and $so(d+1)$
for the Coulomb system.
The hidden symmetry has  a very transparent meaning
in the case of oscillator
 while in  the case of the Coulomb system  it has
a more complicated interpretation
 in terms of geodesic flows of a $d$-dimensional sphere \cite{moser}.
On the other hand, both in classical and quantum cases,
the transformation $r=R^2$ converts
 the $(p+1)-$dimensional radial Coulomb problem
to a  $2p-$dimensional
 radial oscillator  ( $r$ and $R$
denote the radial coordinates of Coulomb and oscillator systems,
 respectively).
In three distinguished cases, $p=1,2,4$,
 one can establish a  complete  correspondence
 between the Coulomb and the oscillator systems,
 by  making use of   the so-called
Bohlin (or Levi-Civita) \cite{bohlin}, Kustaanheimo-Stiefel \cite{ks}
and Hurwitz \cite{h} transformations,
 respectively (see  also \cite{groshe}).
 These transformations  imply
 the reduction of the oscillator system by an  action of $Z_2$,
$U(1)$, $SU(2)$ group, respectively, and yield the
Coulomb-like systems specified by the presence
of  monopoles \cite{ntt,mic,su2}.
%The relation of these transformations with the
%$S^1/Z_2\cong S^1$,
%$S^3/S^1\cong S^2$ and $S^7/S^3\cong S^4$ bundles
%%(and linearizability of $S^1$, $S^3$, $S^7$ spheres)
%causes their applications out of the initial  prescription and include
% twistor description of monopoles, instantons and
%relativistic spinning particles.

On the other hand, the  oscillator and 
Coulomb systems admit the  generalizations to a $d-$di\-men\-si\-onal
sphere and a two-sheet hyperboloid (pseudosphere) with a radius $R_0$
 given by the potentials \cite{sphere,sphere1}
\begin{equation}
V_{osc}=\frac{\alpha^2 R_0^2}{2}\frac{{\bf x}^2}{{x}^2_{d+1}},\quad V_{C}=
-\frac{\gamma}{R_0}\frac{x_{d+1}}{|{\bf x}|},
\label{v}\end{equation}
where ${\bf x}, x_{d+1}$ are the (pseudo)Euclidean coordinates of the ambient
space $\DR^{d+1}$($\DR^{d.1}$):
$\epsilon{\bf x}^2+ x^2_{d+1}=R_0^2$, $\epsilon=\pm 1$.
 The $\epsilon=+1$ corresponds
to the sphere and  $\epsilon=-1$ corresponds to  the pseudosphere.
These systems possess nonlinear hidden symmetries providing
them  with the properties similar to those
 of conventional oscillator and Coulomb systems.
Notice that the oscillators on  sphere and pseudosphere have
 isomorphic  configuration spaces ( $d-$dimensional plane
with a cut circle, in the stereographic projection), since the first one
is undefined on the equator $x_{d+1}=0$.
The Coulomb system is attractive/repulsive on the upper/lower hemisphere
 and has the same behavior on  both the sheets of the hyperboloid.
These systems have been investigated
by various methods  from  many viewpoints (see,  e. g.
\cite{2o} and refs therein).

{\it How to relate the oscillator and Coulomb systems on
(pseudo)spheres ?} \\
This question seems to be crucial for understanding
the geometrical meaning of the hidden symmetries
of Coulomb systems on (pseudo)spheres
and  for the construction of their  generalizations,
as well as for the twistor description of the relativistic spinning
particles on  AdS spaces.
In  Ref.\cite{kmp} devoted to this problem, the oscillator
and Coulomb systems on spheres were related by  mappings
containing  transitions to imaginary coordinates.

In the present letter, we  establish the transparent
 correspondence  between oscillator and Coulomb systems on (pseudo)spheres
for the simplest, two-dimensional, case ($p=1$)
 that can  be extended easily to  higher dimensions ($p=2,4$).
We show that, under stereographic projection  the conventional
Bohlin transformation relates the   two-dimensional oscillator
on the (pseudo)sphere to  the  Coulomb system on pseudosphere,
as well as those interacting with specific external magnetic fields.
This simple  construction allows immediately  to
  connect  the
generators of the hidden symmetries of the systems under consideration,
as well as to clarify the mappings suggested in  \cite{kmp}.

Let us introduce  the complex coordinate
$z$ parametrizing the sphere
by the complex projective plane $\DC P^1$ and
the two-sheeted hyperboloid by  the Poincar\'e  
disks ${\cal L}$
\begin{equation}
{\bf x}\equiv x_1+ix_2=R_0\frac{2z}{1+\epsilon z\bar z},\quad
 x_3=R_0\frac{1-\epsilon {z\bar z}}{1+\epsilon {z\bar z}}.
\label{x}\end{equation}
so that  the metric becomes conformally-flat %takes the K\"ahler form
\begin{equation}
 ds^2=R_0^2\frac{4dz d\bar z}{(1+\epsilon{z\bar z})^2}.%,
\label{met}\end{equation}
%while $R_0x_k$ define the isometries
%of the Kahler structure ($su(2)$ if $\epsilon=1$ and 
%$su(1.1)$ if $\epsilon=-1$).
The lower hemisphere and the lower sheet of the hyperboloid are
parametrized by the unit disk $|z|<1$, while the upper hemisphere
and the upper sheet of hyperboloid are specified by
$|z|>1$, and transform
one into another  by the inversion $z\to 1/z$. 
Since in the limit $R_0\to \infty$  the lower
 hemisphere (the lower sheet of hyperboloid) converts into the
 whole two-dimensional plane, we have to restrict ourselves by those defined
on the lower hemisphere and the 
lower sheet of hyperboloid (pseudosphere),  
to keep  the correspondence with conventional
oscillator and Coulomb systems.
In these terms the oscillator and Coulomb potentials
read
\begin{equation}
V_{osc}=\frac{2\alpha^2 R_0^2z{\bar z}}{(1-\epsilon z{\bar z})^2},
\quad V_{C}=
-\frac{\gamma}{R_0}\frac{1-\epsilon z{\bar z}}{2|z|},
\label{v1}\end{equation}
Notice that  the parametrization (\ref{x})
 is nothing but the stereographic projection
of the two-dimensional (pseudo)sphere 
\begin{equation}
\frac{z}{2R_0}=\left\{
\begin{array}{cc}
\cot\frac{\theta}{2}{\rm e}^{i\varphi}& {\rm for}\;
{\rm sphere};\\
\coth\frac{\theta}{2}{\rm e}^{i\varphi}& {\rm for}\;
{\rm pseudosphere},
\end{array}\right.
\end{equation}
where $\theta,\varphi$ are the (pseudo)spherical coordinates.

Let us equip the  oscillator's phase space
$T^*\DC P^1$ ($T^*{\cal L})$ with the symplectic structure
\begin{equation}
\omega=d\pi\wedge dz+ d{\bar\pi}\wedge d{\bar z}
\label{ss}\end{equation}
and  introduce the rotation generators 
defining  $su(2)$ algebra if $\epsilon=1$
and $su(1.1)$ algebra if $\epsilon=-1$
\begin{equation}
{\bf J}\equiv \frac{iJ_1 -J_2}{2}=\pi +\epsilon{\bar z}^2\bar\pi,
\;\; J\equiv \frac{\epsilon J_3}{2}=i(z\pi-{\bar z}{\bar\pi}).
\label{j}\end{equation}
These generators, together with ${\bf x}/R_0, x_3/R_0$
define the algebra of motion of the (pseudo)sphere
 via  the following  nonvanishing Poisson brackets
\begin{equation}
\begin{array}{c}
\{{\bf J}, {\bf x}\}=2x_3,\;\{{\bf J}, x_3\}=-\epsilon{\bf\bar x},\;
\{J, {\bf x}\}=i{\bf x}, \\
\{{\bf J}, {\bf\bar J}\}=-2i\epsilon J,\;
\{{\bf J}, J\}=i{\bf J}.
\end{array}
\label{alg}\end{equation}
In these terms,  the Hamiltonian of free particle on the (pseudo)sphere
 reads
\begin{equation}
{ H}^{\epsilon}_0=
\frac{{\bf J}{\bf\bar J}+\epsilon J^2}{2R_0^2}
=\frac{(1+\epsilon z{\bar z})^2\pi{\bar\pi}}{2R_0^2}
\label{k}\end{equation}
while the oscillator's Hamiltonian is given
 by the expression
\begin{equation}
{ H}^{\epsilon}_{osc}(\alpha,R_0|\pi, {\bar\pi}, z, {\bar z})=
\frac{(1+\epsilon z{\bar z})^2\pi{\bar\pi}}{2R_0^2}
+\frac{2\alpha^2R_0^2{z\bar z}}{(1-\epsilon{z\bar z})^2}.
\label{ho}\end{equation}
It can be easily verified, by the use of (\ref{alg}), that 
the latter system  possesses the hidden symmetry given by the
complex (or vectorial) constant of motion \cite{sphere1}
\begin{equation}
{\bf I}=I_1+iI_2=\frac{{\bf J}^2}{2R_0^2} +
\frac{\alpha^2R_0^2}{2}\frac{{\bf\bar  x}^2}{x^2_3},
\label{I}\end{equation}
which defines, together  with $J$ and $H_{osc}$ ,  the cubic algebra
\begin{equation}
 \{{\bf I}, J\}=2i{\bf I},\;\;\{{\bf\bar I},{\bf I}\}=4i
\left(\alpha^2 J +\frac{\epsilon JH_{osc}}{R_0^2}-\frac{J^3}{2R_0^4}\right).%\\
\label{Ia}\end{equation}
The energy surface of the oscillator on the (pseudo)sphere
$H^\epsilon_{osc}=E$  reads
\begin{equation}
\frac{\left(1-(z\bar z)^2\right)^2{\pi}{\bar\pi}}{2R_0^4}+
2\left(\alpha^2+\epsilon \frac{E}{R^2_0}\right) z{\bar z}
=\frac{E}{R^2_0}\left(1+(z\bar z)^2\right).
\label{os}\end{equation}
Now, performing the canonical Bohlin transformation \cite{bohlin}
\begin{equation}
w=z^2,\quad p=\frac{\pi}{2z},
\label{boh}\end{equation}
one can rewrite the expression (\ref{os})
as follows:
\begin{equation}
\frac{(1-w\bar w)^2{p}{\bar p}}{2r_0^2}-
\frac{\gamma}{r_0}\frac{1+w{\bar w}}{2|w|}={\cal E}_{C},
\label{C}\end{equation}
where we introduced the notation
\begin{equation}
r_0=R_0^2,\quad  \gamma=\frac{E}{2},
\quad -2{\cal E}_{C}=\alpha^2+\epsilon\frac{E}{r_0}.
\label{gam}\end{equation}
Comparing the l.h.s. of (\ref{C}) with the  (\ref{v1}), (\ref{k})
we  conclude that (\ref{C}) defines the energy surface 
of the Coulomb system 
 on the pseudosphere with ``radius'' $r_0$,
 where  $w, p$ denote complex
 stereographic coordinate and its conjugated momentum, respectively.
 $r_0$ is the ``radius'' of pseudosphere, while ${\cal E}_C$ is the
system's  energy.

{\it Hence, the Bohlin transformation of the
classical isotropic oscillator on the (pseudo)sphere
yields the  classical Coulomb problem on the pseudosphere.} 

The constants of motion of the oscillators, $J$ and ${\bf I}$
(which   coincide
 on the energy surfaces (\ref{os})) are  converted, respectively, 
 into the doubled
angular momentum and the doubled Runge-Lenz vector
 of the Coulomb system
\begin{equation}
J\to 2J_{C},\quad
 {\bf I}\to 2{\bf A},\quad {\bf A}=
-\frac{iJ_{C}{\bf J}_{C}}{r_0}+
{\gamma}\frac{{\bf\bar x}_{C}}{|{\bf x}_{C}|},
\end{equation}
where ${\bf J}_{C}$, $J_{C}$, ${\bf x}_{C}$ denote the rotation generators
and  the pseudo-Euclidean coordinates  of the Coulomb system.

The  quantum-mechanical  counterpart of the energy surface
(\ref{os}) is the Schr\"odinger equation
 \begin{equation}
{\cal H}^{\epsilon}_{osc}(\alpha, R_0| \pi,{\bar\pi},z,{\bar z})\Psi(z,{\bar z})=
E\Psi(z,{\bar z}),
\label{1}\end{equation}
with the quantum Hamiltonian defined
(due to the two-dimensional origin of the system) by the expression
(\ref{ho}), where  $\pi,\bar\pi$  are the momenta operators
(hereafter we assume $\hbar=1$)
\begin{equation}
\pi=-i\frac{\partial}{\partial{z}},\quad
{\bar\pi}=-i\frac{\partial}{\partial{\bar z}} .
\end{equation}
The energy spectrum of this system
is given by the expression (see e.g. \cite{2o} and refs therein)
\begin{equation}
E= {\tilde\alpha}(N+1)+\epsilon\frac{(N+1)^2}{2R^2_0},
\quad N=2n_r+|M|,
\label{oenergy}\end{equation}
where ${\tilde\alpha}=\sqrt{\alpha^2+1/(4R^4_0)}$,
$M$ is the eigenvalue of $J$, $N$ is the principal quantum number,
 $n_r$ is the radial quantum number,
\begin{equation}
\begin{array}{c}
|M|=0,1,\ldots,N,\;\; n_r=0,1,\ldots[ N/2],\cr
N=0,1,\ldots,N_{max}
=\left\{
\begin{array}{cc}
\infty\;, &{\rm if }\; \epsilon=1
\cr
[{2\tilde\alpha}R^2_0]-1\;,&
{\rm if }\; \epsilon=-1
\end{array}
\right.
\end{array}
\end{equation}
So,  the the number of levels in the energy spectrum of the
oscillator is infinite on the sphere and finite on the pseudosphere.
The degeneracy of the energy spectrum  is the same as in
the flat case, viz  $2N+1$.

The quantum-mechanical correspondence
between  oscillator and Coulomb systems is  more complicated,
because the  Bohlin transformation (\ref{boh}) maps
 the $z$-plane in  the
two-sheeted Riemann surface, since ${\rm arg}\;w\in[0,4\pi)$.
 Thus, we have to supplement the quantum-mechanical  Bohlin  transformation
with the reduction by the $Z_2$ group action,
 choosing either
 even  ($\sigma=0$) or odd ($\sigma=1/2$) wave functions
 \begin{equation}
\begin{array}{c}
  \Psi_\sigma (z, \bar z) =
\psi_\sigma (z^2, {\bar z}^2)(z/{\bar z})^{2\sigma}:\\
\psi_{\sigma}(|w|, {\rm arg} w + 2\pi)=
\psi_{\sigma}(|w|, {\rm arg} w ).
\end{array}
\label{3}\end{equation}
 This implies that the range  of
definition of $w$ can be restricted, without loss of generality,
 to ${\rm arg}\; w\in[0,2\pi)$.
In that case, the resulting system is the Coulomb
problem on the hyperboloid given by the Schr\"odinger equation
\begin{equation}
H^{-}_{C}(\gamma, r_0|p_{\sigma}, {\bar p}_\sigma,w,\bar w)
\psi_\sigma={\cal E}_C\psi_{\sigma}
\end{equation}
where   $\gamma, {\cal E}_C, r$ are given by (\ref{gam}),
${H}^{-}_{C}$  denotes the Hamiltonian of the Coulomb system on pseudosphere
with the 
momenta operators 
\begin{equation}
{ p}_\sigma = -i\frac{\partial}{\partial w}-\frac{\sigma}{iw},
\quad{\bar p}_\sigma = -i\frac{\partial}{\partial\bar w}
+\frac{\sigma}{i{\bar w}}.\end{equation}
Hence, the resulting  Coulomb system includes the interaction
 with the   magnetic vortex (an infinitely thin solenoid)
with the magnetic flux $\pi\sigma $  and  zero strength
 $rot\;{\sigma}/{w}=0$. Such  composites are tipical representatives
of  anyonic systems with the spin  $\sigma$ \cite{anyon}.
{\it So, we get  a conventional} $2d$ {\it Coulomb problem
on the hyperboloid at } $\sigma=0$ {\it
  and those with spin $1/2$  generated by the
magnetic flux, at} $\sigma=1/2${\it .}
Taking into account the relations (\ref{gam}),
one can rewrite  the oscillator's energy spectrum (\ref{oenergy})
as follows
\begin{equation}
\sqrt{\frac{1}{4r_0^2}-\epsilon\frac{2\gamma}{r_0}-2{\cal E}_C}=
\frac{2\gamma}{N+1}-\epsilon\frac{N+1}{2r_0}.
\label{inter}\end{equation}
 From this expression one can easily
obtain the   energy spectrum of the reduced system on the pseudosphere
\begin{equation}
{\cal E}_C=-\frac{N_\sigma(N_\sigma+1)}{2r^2_0}-
\frac{\gamma^2}{2(N_\sigma+1/2)^2},
\end{equation}
where
\begin{equation}
\begin{array}{c}
    N_\sigma=n_r+m_\sigma,\quad m_\sigma=M/2, \cr
n_r, m_\sigma-\sigma, N_\sigma-\sigma=0,1,\ldots, N_\sigma^{max}-\sigma.
\end{array}
\end{equation}
Here $m_\sigma$  denotes the eigenvalue of the angular momentum of
the reduced system, and $n_r$ is the radial quantum number of the
initial (and reduced) system.
Notice, that the magnetic vortex shifts the energy levels
 of the two-dimensional Coulomb system
 which is nothing but   the reflection of
Aharonov-Bohm effect.\\
It is seen, that the whole spectrum of the oscillator on pseudosphere
($\epsilon =-1$) transforms into  the spectra of the constructed
Coulomb systems on the pseudosphere, while for the oscillator on
the sphere ($\epsilon =1$) the positivity of
l. h. s. of (\ref{inter})  restricts the admissible values of $N_\sigma$.
 So,  only the part of the spectrum of the oscillator
on the sphere transforms into the spectrum of Coulomb system.
Hence, in  both  cases we get the same  result
 \begin{equation}
 N^{max}_\sigma=\left[\sqrt{r_0\gamma}-(1/2+\sigma)\right].
\end{equation}

To obtain the flat limit  we perform the rescaling
$$(z,\;{\pi})\to(\frac{z}{2R_0},\; 2R_0{\pi}),
\quad (w,\;{p})\to(\frac{w}{4r_0},\; 4r_0{p}),$$
where $r_0=R_0^2$,  and then take the limit $R_0\to\infty$.
In this limit,  the oscillator  on the (pseudo)sphere
results in the conventional circular oscillator
\begin{equation}
    H=2{\pi{\bar\pi}}+\frac{\alpha^2 z{\bar z}}{2},\quad
\omega=d\pi\wedge dz+d{\bar\pi}\wedge d{\bar z},
\end{equation}
which possesses  the hidden $su(2)$ symmetry given by the
 constants of motion
\begin{equation}
\begin{array}{c}
J=i(\pi z-{\bar\pi}{\bar z}),\quad
{\bf I}=2\pi^2+{\alpha^2{\bar z}^2}/{2}:\\
\{{\bf{\bar  I}}, {\bf { I}} \}=4i\alpha^2 J, \quad
\{{\bf I}, J\}=2i{\bf I}.
\end{array}
\end{equation}
The canonical  transformation (\ref{boh})  remains unchanged,
the energy level of oscillator converts into the energy level
of the Coulomb problem with coupling constant $E_{\;}/2$ and the energy
$-\alpha^2/2$. The oscillator's constants of motion $J$ and ${\bf I}$
yield, respectively, the doubled angular momentum and  the doubled
Runge-Lenz vector
$${{\bf A}}=-4ipJ +\frac{E}{2}\frac{\bar{w}}{|w|}.$$
In the quantum case,  the even states of the
oscillator  yield  the conventional Coulomb system,
 while the odd states yield the Coulomb system in the
 presence of magnetic vortex
generating  spin $1/2$\cite{ntt}.

Let us briefly discuss  the Bohlin transformation for the oscillator on
the (pseudo)sphere interacting with constant magnetic field $B_0$.
This system can be defined by the following
replacement of the  symplectic structure   (\ref{ss})
 and of the rotation generators (\ref{j})
\begin{equation}
\omega\to \omega +
{B_0}\frac{i4R_0^2dz\wedge d{\bar z}}{(1+\epsilon z{\bar z})^2},
\quad J_i\to J_i+4R_0B_0{x_i},
\label{r1}\end{equation}
which preserves the algebra (\ref{alg})
and  shifts  the initial Hamiltonian (\ref{ho}) on the constant $(4B_0)^2$.\\
Consequently, the modified system  also possesses 
the hidden symmetry given by (\ref{I}), (\ref{Ia}).
The Bohlin transformation (\ref{boh})
of the modified  oscillator  yields
the Coulomb system on the pseudosphere
interacting with  the  magnetic field
\begin{equation}
B_{C}=\frac{B_0}{2r_0}
\left(\frac{x_{(C)3}}{|{\bf x}|}-\epsilon\right).
\end{equation}

It is easy to see, that the $2p-$dimensional oscillator on (pseudo)sphere
can be  connected to the $(p+1)-$dimensional
Coulomb-like systems on pseudosphere in the same manner  as in  higher
dimensions ($p=2,4$).
Indeed, in stereographic coordinates,
  the
oscillator on $2p$-dimensional
(pseudo)sphere  is described  by the
Hamiltonian  system given by (\ref{ss}), (\ref{ho}), where
 the following replacement is performed
$(z,\;\pi)\to (z^a,\;\pi_a)$, $a=1,\ldots, p$
with the summation over these indices.
Consequently,  the oscillator's energy surfaces  are
of the form (\ref{os}).
Further reduction to the $(p+1)$-dimensional Coulomb-like
system on pseudosphere
repeats the  corresponding reduction in the flat case \cite{mic,su2}.

For example, if $p=2$, we reduce  the system under consideration
by the Hamiltonian action of the $U(1)$ group given by the generator
\begin{equation}
J=i(z\pi-{\bar z}{\bar\pi}).
\end{equation}
For this purpose, we have to  fix the level surface
$J=2s$
and choose the $U(1)$-invariant  stereographic coordinates
in the form of  conventional
Kustaanheimo-Stiefel transformation \cite{ks} (see also \cite{mic})
\begin{equation}
{\bf u}=z{\bs}{\bar z},\quad
{\bf p}=\frac{z{\bs}\pi+{\bar \pi}{\bs}{\bar z}}{2({z\bar z})},
\end{equation}
where $\bs$ are the Pauli matrices. \\
As a result,
the reduced
symplectic structure reads
\begin{equation}
d{\bf u}\wedge d{\bf p} +
s\frac{({\bf u}\times d{\bf u})\wedge d{\bf u}}{|{\bf u}|^3},
\label{ss2}\end{equation}
while the oscillator's energy surface
takes the form
\begin{equation}
\frac{(1- {\bf u}^2)^2}{8r_0^2}({\bf p}^2 +\frac{s^2}{{\bf u}^2})-
\frac{\gamma}{r_0}\frac{1+{\bf u}^2}{2|{\bf u}|}={\cal E}_C,
\label{C3}\end{equation}
where  $r_0$, $\gamma$, ${\cal E}_C$
are defined by the expressions (\ref{gam}).

Interpreting ${\bf u}$  as the
 real stereographic coordinates of three-dimensional
pseudosphere 
\begin{equation}
{\bf x}=r_0\frac{2{\bf u}}{1- {\bf u}^2},\quad
 x_4=r_0\frac{1+ {\bf u}^2}{1- {\bf u}^2},
\label{hyper3}
\end{equation}
  we conclude that (\ref{C3}) defines  the energy surface of 
the  pseudospherical analog of a Coulomb-like system 
proposed in Ref.\cite{z}, that  describing the interaction of two 
non-relativistic dyons.

In $p=4$ case, we have to reduce the system by  the  action
of the $SU(2)$ group
and choose the $SU(2)-$invariant stereographic coordinates and momenta
in the form  corresponding to the standard Hurwitz transformation
\cite{h,su2}
which yields a pseudospherical analog of the so-called $SU(2)$-Kepler 
(or Yang-Coulomb) system  \cite{su2}.\\

{\large Acknowledgments.} The authors are grateful  to
V. M. Ter-Antonyan for valuable discussions
and C. Groshe for drawing their  attention to Ref.\cite{groshe}.
 A.N. thanks D. Fursaev and  C. Sochichiu  for useful comments and interest 
in the work.
The work of G.P. is partially supported by RFBR grants 98-01-00330 and
00-02-81023.

\end{document}